\documentclass{jpsj3}
\usepackage{graphicx,amsmath,amssymb,bm}

\allowdisplaybreaks 

\usepackage{color}
\definecolor{Green}{rgb}{0,0.7,0}

\begin{document}
\title{
Berry Curvature  of the Dirac Particle   
in  $\alpha$-(BEDT-TTF)$_2$I$_3$ 
\;\;\; 
}
\author{
Yoshikazu  \textsc{Suzumura}$^{1}$ and 
Akito  \textsc{Kobayashi}$^{1,2}$,
}

\inst{
$^1$Department of Physics, Nagoya University, 
 Chikusa-ku, Nagoya, 464-8602 Japan \\
$^2$Institute for Advanced Research, Nagoya University, 
 Chikusa-ku, Nagoya, 464-8602 Japan \\
}

\abst{
We examine several properties of the Berry curvature for the organic conductor $\alpha$-(BEDT-TTF)$_2$I$_3$ consisting of four bands, which exhibits a zero-gap state with  Dirac cones. 
By adding a small potential acting on two molecular sites, which breaks the inversion symmetry, it is shown that the curvature for the Dirac particles displays a pair of peaks with opposite signs 
 and that each peak increases with decreasing potential. 
The Berry curvature  originating from  the property of the wave function is analyzed using a reduced Hamiltonian with a  2x2 matrix based on the 
 Luttinger-Kohn representation, 
 which describes a pair of Dirac particles between the conduction band and the valence band. Two types of velocity fields in the reduced Hamiltonian, whose vector product gives the Berry  curvature,  rotate around the Dirac point as a vortex. It is also shown that the other bands exhibit another pair of peaks of  Dirac particles with   a tendency toward merging. 
}

\kword{
Berry curvature,
Dirac particles,
Dirac point, 
zero gap,
Luttinger-Kohn representation,
$\alpha$-(BEDT-TTF)$_2$I$_3$,
organic conductor
}

\maketitle

\section{Introduction}
The metal-insulator transition 
 in the two-dimensional organic conductor $\alpha$-(BEDT-TTF)$_2$I$_3$ 
\cite{Mori1984} (BEDT-TTF=bis(ethylenedithio)tetrathiafulvalene)
 has been studied extensively since 
   the discovery of electric conductivity, which  
 exhibits a noticeable dependence  on both pressure and magnetic field.
\cite{Kajita1992}
With increasing uniaxial pressure $P_a$  along the stacking axis, the insulating state changes into  
the superconducting state, and is followed 
  by a narrow-gap semiconducting state.\cite{Tajima2002}
  The  Hall conductivity at  $P_a =$ 10 kbar decreases rapidly with decreasing temperature.\cite{Tajima2002} Using the transfer energy obtained from the structure analysis,\cite{Kondo2005}  the successive transition has been studied theoretically in terms of an extended Hubbard model.\cite{Kobayashi2005JPSJ} 
 In addition to  the vanishment of the density of states  
    in the narrow-gap state,\cite{Kobayashi2004JPSJ}  
    it has been demonstrated that the zero-gap state exhibits a contact point 
followed by a pair of Dirac cones.\cite{Katayama2006JPSJ} 
The zero-gap state was confirmed 
   by first-principles calculation.\cite{Kino2006,Ishibashi2006}
 Such a Dirac particle,  can explain the anomalous temperature dependence of resistivity\cite{Tajima2007EPL} and Hall coefficient.\cite{Kobayashi2008JPSJ}
\begin{figure}
  \centering
\includegraphics[width=8cm]{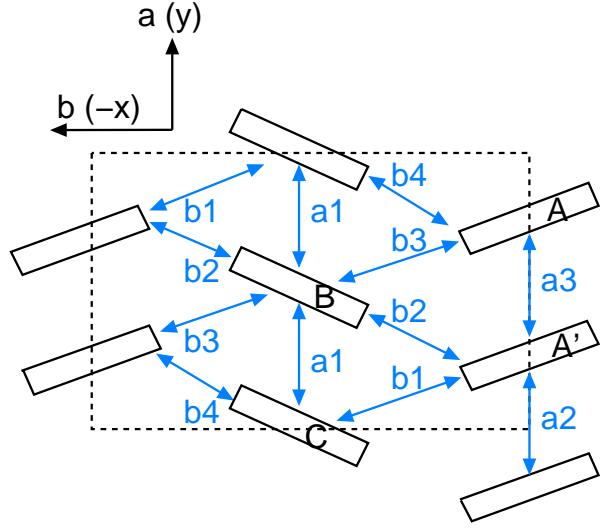}   
  \caption{(Color online)
Crystal structure on two-dimensional plane with four molecules 
 A(1), A'(2), B(3), and C(4), 
in the unit cell 
where the respective bonds represent 
 seven transfer energies 
$t_{b1}, \cdots, t_{a3}$.
}
\label{fig:structure}
\end{figure}

 In Fig.~\ref{fig:structure}, the crystal structure of
  $\alpha$-(BEDT-TTF)$_2$I$_3$  is shown where there are   
 four sites, i.e.,  A, A', B, and C,  in the unit cell
 with an inversion symmetry between the A and A' sites. 
 For the insulating state, which is  found  at ambient pressure, 
  the symmetry is broken since the electron density at the A site is different 
      from that at the  A' site owing to 
     a  stripe charge ordering perpendicular to the stacking axis.
\cite{SeoRev2004,Seo2000,Kino1995,Takahashi2003} 
At high pressures  where the zero-gap state emerges, 
 the symmetry is restored, and the electron density  at the A site becomes equal to that at the A' site. \cite{Kakiuchi2007}
The zero-gap state is robust to the variation in pressure, 
  although  the contact point   
     moves in the Brillouin zone (BZ).\cite{Katayama2006JPSJ} 
Small amounts of anion potential which are the same for the A and A' sites 
 but are different for the B and C sites, do not break the contact point of the Dirac cone.
\cite{Katayama2009EPJB}
These site potentials for B and C 
 also stabilize  the zero-gap state  by reducing the overlap between 
 the conduction  and  valence bands.   
\cite{Kondo2009}
A pair of Dirac cones located at the incommensurate wave vector  
   in the BZ is expected to vanish at high pressures 
by merging at a symmetric point (e.g., $\Gamma$ point).
\cite{Kobayashi2007JPSJ,Montambaux2009EPJB,Montambaux2009PRB}
 Furthermore, it has been shown that the four sites, A, A', B, and C 
  provide the respective role for the tilted Dirac cone, and that 
 each site gives a difference in the local density of states, 
as shown in the temperature dependence of the Knight shift and 
$T_1^{-1}$ of NMR.
\cite{Katayama2009EPJB,Takahashi2010}

 Such a Dirac particle can be represented in terms of a reduced Hamiltonian 
  with a 2x2 matrix instead of the original 4x4 matrix Hamiltonian, 
 where the reduced Hamiltonian based on 
      the Luttinger-Kohn (LK) representation,
\cite{Luttinger1955} gives 
    the tilted Weyl equation 
      \cite{Kobayashi2007JPSJ,Kobayashi2008JPSJ,Goerbig2008PRB}
         for the Dirac particle.  
 The reduced model is successful to examine  
 the novel property of transport phenomena, 
\cite{Kobayashi2008JPSJ,Kobayashi2009JPSJ,Nishine2010}
 which  exhibit 
 several differences  from those of  graphene, 
\cite{Ando2005JPSJ,Novoselov2005}
owing  to tilting. 

Although much progress has been achieved on the basis of the zero-gap state
,
\cite{Tajima2006JPSJ,Tajima2009STAM,Kobayashi2009STAM} 
 the condition for the Dirac particle is not yet clear since a contact point 
 is formed owing to the accidental degeneracy in the energy band.
\cite{Herring1937}
In the present study, 
as the next step in understanding  the Dirac particle, 
we examine the Berry curvature\cite{Berry1984,Aharonov1987PRL,Xiao} 
 for the  $\alpha$-(BEDT-TTF)$_2$I$_3$ salt, 
 which comes from the property of the wave function.
 The curvature is also a useful tool for 
  searching for  the  particle in  other organic conductors. 
  The Dirac particle has  been  maintained 
 using a fact that  the gap  between the first band and the second band is zero within  numerically accuracy. 
  However, this is not always the case for the Dirac particle since there is also a Dirac particle with a gap,
e.g.,   a massive Dirac particle in the low-energy model of boron nitride,\cite{Fuchs2010}  and also  in the charge ordered state of $\alpha$-(BEDT-TTF)$_2$I$_3$ at pressures  slightly lower than those of the zero-gap state.
\cite{Kobayashi2010Merging}

In \S 2, the formulation for the Berry curvature of $\alpha$-(BEDT-TTF)$_2$I$_3$
  is given using a 4x4 Hamiltonian. 
  To calculate the Berry curvature 
   with a finite magnitude around the contact point, 
    we  add a small potential that differs between  
   the  A and A' sites,
   based on the previous work.
\cite{Kobayashi2009JP} 
Furthermore,  a reduced Hamiltonian of a 2x2 matrix is examined  by calculating  
 the analytical property  of the curvature. 
In \S3,  the Berry curvatures of the Dirac particle with multibands are studied numerically  by taking a typical pressure,  $P_a$ =6 kbar. 
Velocity fields and Berry curvature are also  studied.  The summary and discussion are given in \S 4.

\section{Formulation and Hamiltonian}
\subsection{Berry curvature of multibands}

We consider a two-dimensional electron system consisting of  $M$ molecular sites in a unit cell where $M=4$ in the present case.
The Schr\"{o}dinger equation is given by 
\begin{align} 
 H(\bm{k}) |n(\bm{k})> = E_n |n(\bm{k})> \; ,
\end{align}
 where  $\bm{k} =(k_x,k_y)$ is  a  two-dimensional  wave vector, and
 $n (= 1, 2, \cdots, M)$ denotes a band index. The quantities 
 $H(\bm{k})$,  $E_n(\bm{k})$, and  $|n(\bm{k})>$  are the Hamiltonian with 
 an MxM matrix for the $M$ sites,   eigenvalue (band energy),  and  eigenfunction (wave function), respectively. 
When  $\bm{k}(t)$ varies slowly with time $t$, the equation for motion is written as   
\begin{align} 
i \hbar \partial_t|n(\bm{k}(t))> = H(\bm{k}(t)) |n(\bm{k}(t))> .
\end{align}
When $\bm{k}$ moves along a closed loop C, the state $|n(\bm{k})>$ after returning to the original location is  replaced by ${\rm e}^{i \gamma_n} |n(\bm{k})>$. The quantity $\gamma_n$ is  the Berry phase  given by
\cite{Berry1984} 
 \begin{align}
  \gamma_n = & {\rm i} \int_{\rm C} 
  \left< n(\bm{k}) \cdot \bm{\nabla}_{\bm{k}} n(\bm{k}) \right> {\rm d} \bm{k}
 =  \int_{\rm S} \bm{B}_n(\bm{k}) \cdot {\rm d} \bm{S} ,
\label{eq:Berry_C}
\end{align}
 where S is  a region enclosed by the loop C.
 The vector field  $\bm{B}_n(\bm{k})$ is the  Berry curvature expressed as   
\begin{align} 
 \bm{B}_n(\bm{k})
  & =  -  {\rm Im} \left<\bm{\nabla}_{\bm{k}} n |  \times 
 | \bm{\nabla}_{\bm{k}} n \right> , 
\label{eq:formula_1}
 \end{align} 
 which is oriented perpendicularly to the $k_x$-$k_y$ plane. 
 Noted that $\bm{B}_n = 0$ for a single component owing to $<n|n>=1$.
 
The  Berry curvature can be rewritten as 
\cite{Berry1984} 
\begin{align} 
 \bm{B}_n (\bm{k}) & 
 = -  {\rm Im} \sum_{m (\not= n)} 
\frac{
 \left<n| \bm{\nabla}_{\bm{k}} H | m \right>  \times 
 \left< m| \bm{\nabla}_{\bm{k}} H | n \right>
 }{ (E_m - E_n)^2} \; ,
\label{eq:formula_2}
 \end{align} 
which is used for the numerical calculation. Note that   
\begin{align} 
\sum_{n=1}^4  \bm{B}_n (\bm{k}) = 0  \; ,
\end{align}
for an arbitrary $\bm{k}$ since $\left<n| \bm{\nabla}_{\bm{k}} H | m \right>$ is  complex conjugate to  $\left< m| \bm{\nabla}_{\bm{k}} H | n \right>$, and the double summations of $n$ and $m$ result in the vanishment of the imaginary part.  
This suggests a possible  pair of  Dirac cones located  between two neighboring energy bands, because the singularity of one band must be compensated with that of  the other band. 

\subsection{Hamiltonian for $\alpha$-(BEDT-TTF)$_2$I$_3$ }
The contact point, which is obtained in a normal state, i.e.,  without charge ordering, is determined essentially by the property of the transfer energy, and  
the effect of interaction is to modify mainly the location of the contact point (see Fig.~5 in ref. \citeonline{Kobayashi2007JPSJ}).
Then, by taking account of only transfer energy, we examine   
the Hamiltonian for $\alpha$-(BEDT-TTF)$_2$I$_3$ given by  
\begin{align}
\tilde{H}= \sum_{\alpha,\beta=1}^4 
   \sum_{\sigma} \sum_{\bm{i},\bm{j}}^{\rm N} 
 t_{\alpha,\beta;\bm{i},\bm{j}} a_{\alpha\sigma,\bm{i}}^{\dagger}
   a_{\beta\sigma,\bm{j}} \; ,
  \label{eq:transfer}
  \end{align}
where 
 $\alpha, \beta$ (=1, 2, 3,4)   are indices 
   for the sites of molecules A (1), A'(2), B (3), and C (4) 
    in the unit cell, and $\bm{i}, \bm{j}( = \bm{R}_i, \bm{R}_j)$ are 
  those  for the cell forming a square lattice with N sites. 
 The quantity $a_{\alpha,\bm{j}}^{\dagger}$ denotes the creation operator
 for the electron, and  $t_{\alpha,\beta;\bm{i},\bm{j}}$ denotes 
  the transfer energy between the neighboring site. 
As shown in Fig.~\ref{fig:structure}, 
 there are seven transfer energies given by 
   $t_{ b1} \cdots t_{b4}$  for the direction of  the $b(-x)$-axis
 and  by $t_{a1} \cdots t_{a3}$   for the direction of the $a(y)$-axis.
We note that the definition of transfer energies 
 in Fig.~\ref{fig:structure} (ref.~\citeonline{Mori1984}) 
 differs from our previous one,
 $t_{p1}, \cdots, t_{c4}$.\cite{Katayama2006JPSJ}
 Since the two-dimensional plane of the previous one
 is reversed compared with that in Fig.~\ref{fig:structure},   
  the relation is given by 
$t_{p1}=t_{b2}$, $t_{p2}=t_{b1}$, $t_{p3}=t_{b4}$, $t_{p4}=t_{b3}$,   
 $t_{c1}=t_{a2}$, $t_{c2}=t_{a3}$, and $t_{c3} (=t_{c4})= t_{a1}$. 
 In order to calculate the Berry curvature, which 
  becomes finite around the Dirac point,
   we introduce  site  potentials $I_{\alpha \sigma}$ 
   acting on A (1)  and A'(2)  differently in the form of 
\begin{align}
\tilde{H}_{\rm site}
 = \sum_{\alpha,\beta=1}^4 \sum_{\bm{i},\bm{j}}^{\rm N} 
I_{\alpha \sigma} \delta_{\bm{i},\bm{j}}
 \delta_{\alpha, \beta}
 a_{\alpha,\sigma,\bm{i}}^{\dagger}a_{\beta,\sigma,\bm{j}} \; ,
 \label{eq:site_pot} 
\end{align}
where 
\begin{eqnarray}
 I_{1\sigma}= - I_{2\sigma} = - \Delta_0, 
\end{eqnarray}
and $I_{3\sigma} = I_{4\sigma} =0$. 
Equation (\ref{eq:site_pot}) is  a Hamiltonian of the site potential 
 that creates a gap 
 by  breaking the inversion symmetry between A and A'.  

Using the Fourier transform 
$ a_{\alpha}(\bm{k}) = \frac{1}{\sqrt{N}}
    \sum_{\bm{j}}  a_{\alpha,\bm{j}} \exp [ - i \bm{k} \cdot \bm{R}_{j}]
$, 
the total Hamiltonian is expressed as 

\begin{eqnarray}
\tilde{H} + \tilde{H}_{\rm site}
& = &
\sum_{\bm{k}\sigma}
\begin{pmatrix} 
a_{1\sigma}(\bm{k})^\dagger,&a_{2\sigma}(\bm{k})^\dagger,& a_{3\sigma}(\bm{k})^\dagger,&a_{4\sigma}(\bm{k})^\dagger
\end{pmatrix} 
H(\bm{k})
\begin{pmatrix}
a_{1\sigma}(\bm{k})\\
a_{2\sigma}(\bm{k})\\
a_{3\sigma}(\bm{k})\\
a_{4\sigma}(\bm{k})
\end{pmatrix} , 
  \\
H({\bf k}) &=  &
\begin{pmatrix}
  - \Delta_0   
  &  t_{c1} + t_{c2} {\rm e}^{- i k_y} 
  &  t_{p1} - t_{p4} {\rm e}^{i k_x} 
  & t_{p2} -t_{p3} {\rm e}^{i k_x}           \\
  t_{c1} + t_{c2} {\rm e}^{ i k_y} 
  &  \Delta_0  
  &  t_{p4}{\rm e}^{i k_y} - t_{p1} {\rm e }^{i (k_x + k_y)}
  &  t_{p3} - t_{p2} {\rm e}^{i k_x}   \\
  t_{p1} - t_{p4} {\rm e}^{- i k_x} 
  &  t_{p4}{\rm e}^{-i k_y} - t_{p1} {\rm e }^{-i (k_x + k_y)}
  & 0 
  & t_{c3} + t_{c4} {\rm e}^{- i k_y} \\
  t_{p2} -t_{p3} {\rm e}^{- i k_x}  
 & t_{p3} - t_{p2} {\rm e}^{- i k_x}  
 &  t_{c3} + t_{c4} {\rm e}^{ i k_y} 
 & 0
\end{pmatrix}
 \nonumber \\
.
\label{eq:CO_Hamiltonian}
\end{eqnarray} 
The lattice constant is taken as unity. 
By diagonalizing eq.~(\ref{eq:CO_Hamiltonian}), we obtain four bands 
 $E_j({\bf k})$ $(j=1, \cdots,4)$, where 
 $E_1({\bf k}) >  E_2({\bf k}) > E_3({\bf k}) > E_4({\bf k})$. 
  For  $I_{\alpha \sigma} = 0$
    where the inversion symmetry between A and A' is maintained, 
  the zero-gap state between   $E_1({\bf k})$ and  $E_2({\bf k})$
 is obtained at $\pm \bm{k}_0$, i.e., 
  with the uniaxial pressure along the $a$-axis, 
 $P_a$, being larger than 3 kbar.\cite{Katayama2006JPSJ}
 In the present paper, 
  we introduce a small but finite $\Delta_0$ 
   that induces a small gap 2$\Delta$ between two upper bands. 
We call $\bm{k}_0$ the Dirac point where 
  $\bm{k}_0$ for $\Delta_0=0$ gives  the contact point, and  
  $\bm{k}_0$ for $\Delta_0 \not= 0$ corresponds to  the minimum of the gap between 
 $E_1(\bm{k})$ and  $E_2(\bm{k})$.

\subsection{Reduced Hamiltonian} 
 We  analytically examine  $\bm{B}_1$, which denotes the curvature for  
 the first band $E_1$.
The present numerical calculation shows
  that   $\bm{B}_1$ in eq. (\ref{eq:formula_2}) is essentially 
   determined  by $E_1$ and $E_2$, i.e., 
     the effects of $E_3$ and $E_4$ on $\bm{B}_1$ are negligibly small.   
 Thus, eq.~(\ref{eq:formula_2}) may be  rewritten as 
\begin{align} 
 \bm{B}_1 (\bm{k}) 
= & -  {\rm Im} 
\frac{
 \left<1(\bm{k})| \bm{\nabla}_{\bm{k}} H | 2(\bm{k}) \right>  \times 
 \left< 2(\bm{k})| \bm{\nabla}_{\bm{k}} H | 1(\bm{k}) \right>
 }{ (E_2 - E_1)^2}
 \nonumber \\ 
 &
 = \frac{2}{(E_2 - E_1)^2}(\bm{v}_1(\bm{k}) \times \bm{v}_2(\bm{k})) ,
\label{eq:formula_reduced3}
 \end{align} 
where $\bm{v}_1$ and $\bm{v}_2$ are the  velocity fields defined by 
\begin{align} 
 \left< 1(\bm{k})| \bm{\nabla}_{\bm{k}} H | 2(\bm{k}) \right>
 = \bm{v}_1(\bm{k})  + {\rm i} \bm{v}_2(\bm{k}) \; . 
\label{eq:def_velocity}
  \end{align} 
Although the velocities $\bm{v}_1$ and $\bm{v}_2$ depend
  on the choice of the relative phase between 
$|1(\bm{k})>$ and  $|2(\bm{k})>$, 
 the product of $\bm{v}_1 \times \bm{v}_2$ is
 independent of the phase.

From eq.~(\ref{eq:def_velocity}), one can reduce  
  the 4x4 Hamiltonian  to the 2x2 Hamiltonian
 for the energy bands $E_1$ and $E_2$. 
Instead of eq.~(\ref{eq:formula_reduced3})
 where the band basis on $\bm{k}$ is taken, 
  the effective Hamiltonian is expressed 
    using the LK basis, which is fixed at
 $\bm{k}^{\rm LK} (= \bm{k}_0)$.
The general form of the 2x2 Hamiltonian,  $H^{\rm red} (\bm{k})$,
  is given by 
 \begin{eqnarray}
 H^{\rm red} (\bm{k}) =  H^{\rm LK}(\bm{k})  & = &
 \begin{pmatrix}
f_3 + f_0 & f_1-if_2 \\
f_1+if_2 & - f_3 + f_0
\end{pmatrix}
 \nonumber \\ 
 & &
 = f_0 + f_1 \sigma_1 + f_2 \sigma_2 + f_3 \sigma_3 \; .
\label{eq:Heff_start}
  \nonumber \\
\end{eqnarray}
The quantity $\sigma_j$ denotes the Pauli matrix and 
 the coefficient $f_j$ (= $f_j(\bm{k})$) is given as the function of 
  the two-dimensional wave vector $\bm{k}$. 
 Although the term $f_0$ gives the effect of the tilting for  the Dirac cone,
\cite{Kobayashi2009JPSJ}
  $f_0$ is irrelevant to $\bm{B}$ and is discarded hereafter. 
Substituting eq.~(\ref{eq:Heff_start}) into eq.~(\ref{eq:formula_reduced3}),
we obtain the Berry curvature as (see Appendix)
\begin{align}
\bm{B} = - \frac{1}{2E^3} \left(
 f_3(\bm{\nabla}_{\bm{k}}f_1 \times \bm{\nabla}_{\bm{k}}f_2)
 +   f_1(\bm{\nabla}_{\bm{k}}f_2 \times \bm{\nabla}_{\bm{k}}f_3)
 \right. 
  \nonumber \\
  \left.
 +   f_2(\bm{\nabla}_{\bm{k}}f_3 \times \bm{\nabla}_{\bm{k}}f_1)
\right) ,
\label{eq:Berry_curvarure0}
  \end{align}  
where $E=\sqrt{f_1^2+f_2^2+f_3^2}$.

 Now, we examine the Berry curvature close to the Dirac point  $\bm{k}_0$ 
 using the LK basis at $\bm{k}^{\rm LK}$.
 The quantity   $f_j(\bm{k})$ is expanded as  
 \begin{align}  
& f_1(\bm{k}) 
    \simeq  \bm{v}^0_1 \cdot \bm{q} , 
  \nonumber \\
& f_2(\bm{k})    \simeq \bm{v}^0_2 \cdot \bm{q}, 
\nonumber \\
& f_3(\bm{k}) \simeq -  \Delta ,
\label{eq:eq3-5}
 \end{align}   
where $\bm{q} = \bm{k}-\bm{k}^{\rm LK}$,
  $\bm{v}^0_j ( = \bm{\nabla}_{\bm{k}}f_j) $
 is estimated at  $\bm{k} = \bm{k}^{\rm LK}$, 
 and the linear term in $f_3$ is absent.
The quantity 
 $2 \Delta (< 2 \Delta_0)$,
   which is the gap between 
 the two bands of $E_1$ and $E_2$, is 
 small compared with  the bandwidth. 
 From eqs.~(\ref{eq:Berry_curvarure0}) and (\ref{eq:eq3-5}), 
 the  Berry curvature and Berry phase are respectively obtained as
\begin{align}
\bm{B} =  \frac{\Delta}{2E^3} \left(
\bm{v}^0_1 \times \bm{v}^0_2
 \right) ,
\label{eq:Berry_Phase1}
\end{align} 
\begin{align}
 \gamma_1 = \int ( \bm{e}_z \cdot\bm{B}) {\rm d} \; \bm{q} 
  =  \pi \frac{ \bm{e}_z \cdot(\bm{v}^0_1 \times \bm{v}^0_2)}
{| \bm{e}_z \cdot (\bm{v}^0_1 \times \bm{v}^0_2)|},
\label{eq:eq28}
\end{align} 
 where  $E = \sqrt{f_1^2 + f_2^2 + \Delta^2}$ and  $\bm{e}_z$ denotes the unit vector perpendicular to the $q_x$-$q_y$ plane. 
In deriving eq.~(\ref{eq:eq28}), 
 the two-dimensional integral is extended to infinity. 

Now, we examine the variation in the velocity 
 fields around the Dirac point  
 for an arbitrary $\bm{k} (\not= \bm{k}^{\rm LK}$)
 using eq.~(\ref{eq:def_velocity}). 
 By considering the case of  
 \begin{align}
\bm{v}^0_1 \rightarrow  (v,0),\;\;\;  \bm{v}^0_2 \rightarrow  (0,v') \;\;\; 
  {\rm and}\;\;\;  f_3 \rightarrow - \Delta ,
 \label{eq:special_case}
 \end{align}
we obtain  (see eqs.~(\ref{eq:v1_eff}) and (\ref{eq:v2_eff}) in Appendix)  
 \begin{align}
 & \bm{v}_1(\theta_{\bm{q}}) =  \frac{\Delta}{E \sqrt{E^2 -\Delta^2}}
  (v^2q_x \bm{e}_x + (v')^2q_y \bm{e}_y) \; ,
\label{eq:eq41}
 \\ 
& \bm{v}_2(\theta_{\bm{q}}) =  \frac{1}{ \sqrt{E^2 -\Delta^2}}
  vv'( - q_y \bm{e}_x + q_x \bm{e}_y)
 \; ,
\label{eq:eq42}
\\
& \bm{B} = \frac{\Delta v v'}{2E^3} \bm{e}_z \;,
\label{eq:eq43}
 \end{align}
 where $E = \sqrt{v^2q_x^2+ v'^2q_y^2 + \Delta^2}$.
These  velocity fields rotate  around $\bm{k}_0$
 as shown explicitly for the 4x4 Hamiltonian in the next section.  

 Here, we comment on the 
    reduced Hamiltonian of eq.~(\ref{eq:special_case})  rewritten as 
\begin{eqnarray}
H^{(I)} 
  =&
\begin{pmatrix}
-\Delta & q_x-iq_y \\
q_x+iq_y & \Delta
\end{pmatrix} 
 =  q_x \sigma_1 + q_y \sigma_2 - \Delta \sigma_3 \; ,
  \nonumber \\
\label{eq:Koba_KT}
\end{eqnarray}
 where $v=v'=1$. 
By introducing a unitary transformation,  
 eq.~(\ref{eq:Koba_KT})   is  transformed as follows. 
The transformation for ${\rm e}^{- i \sigma_2 \phi/2}$ with $\phi=\pi/2$
 gives  
\begin{eqnarray}
H^{(II)}= &{\rm e}^{- i \sigma_2 \phi/2} H^{(I)} {\rm e}^{i \sigma_2 \phi/2}
 \nonumber \\
&
 = 
\begin{pmatrix}
q_x & \Delta -iq_y \\
\Delta +iq_y & - q_x 
\end{pmatrix} 
 =  \Delta \sigma_1 + q_y \sigma_2 + q_x \sigma_3 \; , 
  \nonumber \\
\label{H:Kata}
\end{eqnarray}
 and that for ${\rm e}^{ i \sigma_1 \phi/2}$ 
 with  $\phi = \pi/2$ gives 
\begin{eqnarray}
H^{(III)}=& {\rm e}^{ i \sigma_1 \phi/2} H^{(I)} {\rm e}^{-i \sigma_1 \phi/2}
 \nonumber \\
&
 = 
\begin{pmatrix}
k_y & k_x -i\Delta \\
k_x +i\Delta & - k_y 
\end{pmatrix} 
 =  k_x \sigma_1 + \Delta \sigma_2 + k_y \sigma_3 \; .  
  \nonumber \\
\label{H:Koba_Hall}
\end{eqnarray}
 For $\Delta =0$,   eqs.~(\ref{H:Kata}) and  (\ref{H:Koba_Hall}) 
  are reduced to those  derived by Katayama et al.\cite{Katayama2009EPJB} 
    and Kobayashi et al.\cite{Kobayashi2008JPSJ}, respectively, 
    who chose different  bases for the L-K state with   
        $\bm{v}^0_1$ and $\bm{v}^0_2$ being orthogonal.
 Note that all these cases lead to $B_z = \bm{B}\cdot\bm{e}_z =  \Delta/(2E^3)$
 owing to the rotation that does not depend  on $\bm{q}$.

\section{Berry Curvature for $\alpha$-(BEDT-TTF)$_2$I$_3$ }

To calculate the Berry phase of the conductor  
 under uniaxial  $P_a$ along the stacking axis, 
we use an extrapolation\cite{Kobayashi2005JPSJ} 
 for   $t_{\alpha \beta}$  given by
\cite{Kondo2005}  
\begin{eqnarray} 
  t_{\alpha \beta}(P_a) = t_{\alpha \beta}(0) +  C_{\alpha \beta} P_a \; .
\label{eq:extraporation}
\end{eqnarray} 
The unit of  energy is taken as eV and that of 
 pressure $P_a$ is taken as kbar. 
Equation (\ref{eq:extraporation}) corresponding to 
  $\alpha$-(BEDT-TTF)$_2$I$_3$ gives the zero-gap state for $P_a > 3$ kbar.
We perform the  calculation by taking  $P_a$ = 6 kbar and $\Delta_0=$ 0.02 eV.
 The gap on the Dirac point ${\bf k}_0$, i.e.,  $2\Delta (\simeq$ 0.02 eV), 
  is smaller than  $2\Delta_0$.
\begin{figure}
  \centering
\includegraphics[width=6cm]{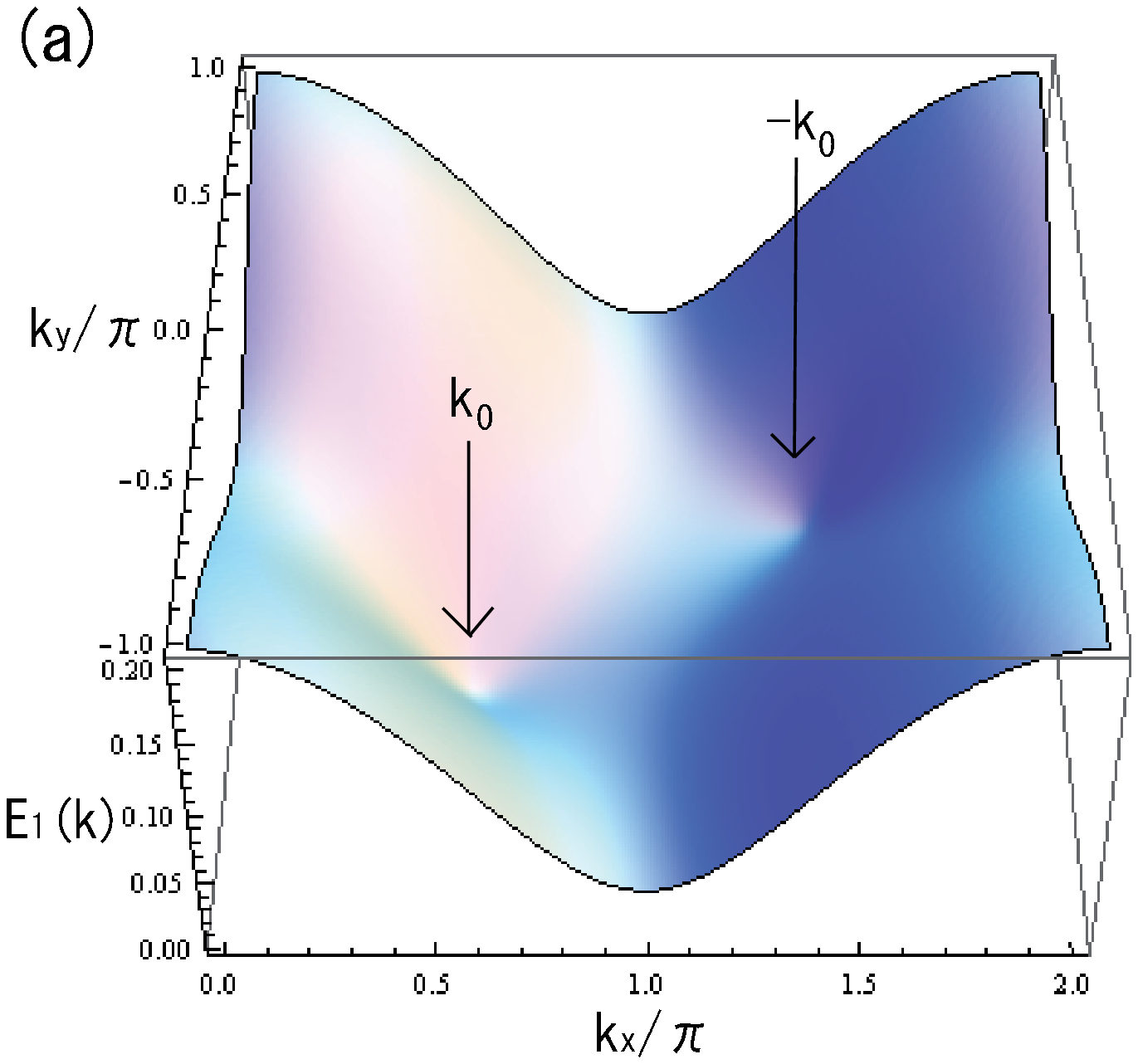}   
\includegraphics[width=6cm]{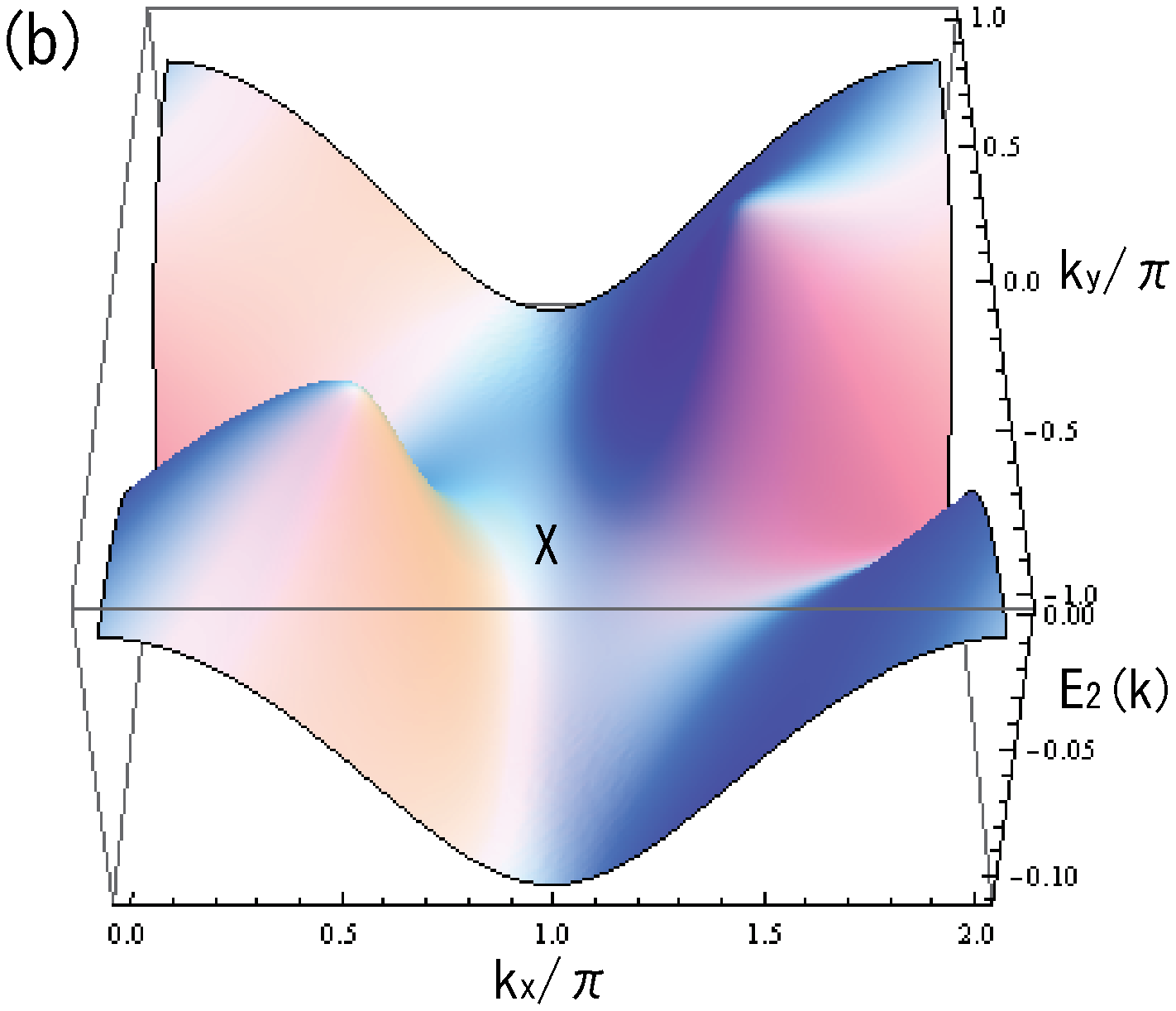}   
  \caption{(Color online) 
Energy bands of $E_1(\bm{k})$ (a) and   $E_2(\bm{k})$ (b)
  for $P_a =6$ kbar and $\Delta_0 = $0.02 eV,
 where the regions for the $k_x$ and $k_y$ axes are given 
        as $0 < k_x < 2\pi$ and $-\pi < k_y < \pi$. 
The arrow denotes the location for the Dirac point 
    at ${\bf k}_0 = (0.57 \pi, - 0.30\pi)$ where 
      the gap between  $E_1(\bm{k})$  and   $E_2(\bm{k})$ is estimated as  
       $2\Delta \simeq$ 0.02 eV. 
Two cones are situated symmetrically with respect to the X point $(\pi,0)$ 
 and the  $\Gamma$ point $(0,0)$. 
}
\label{fig:fig1}
\end{figure}
\begin{figure}
  \centering
\includegraphics[width=6cm]{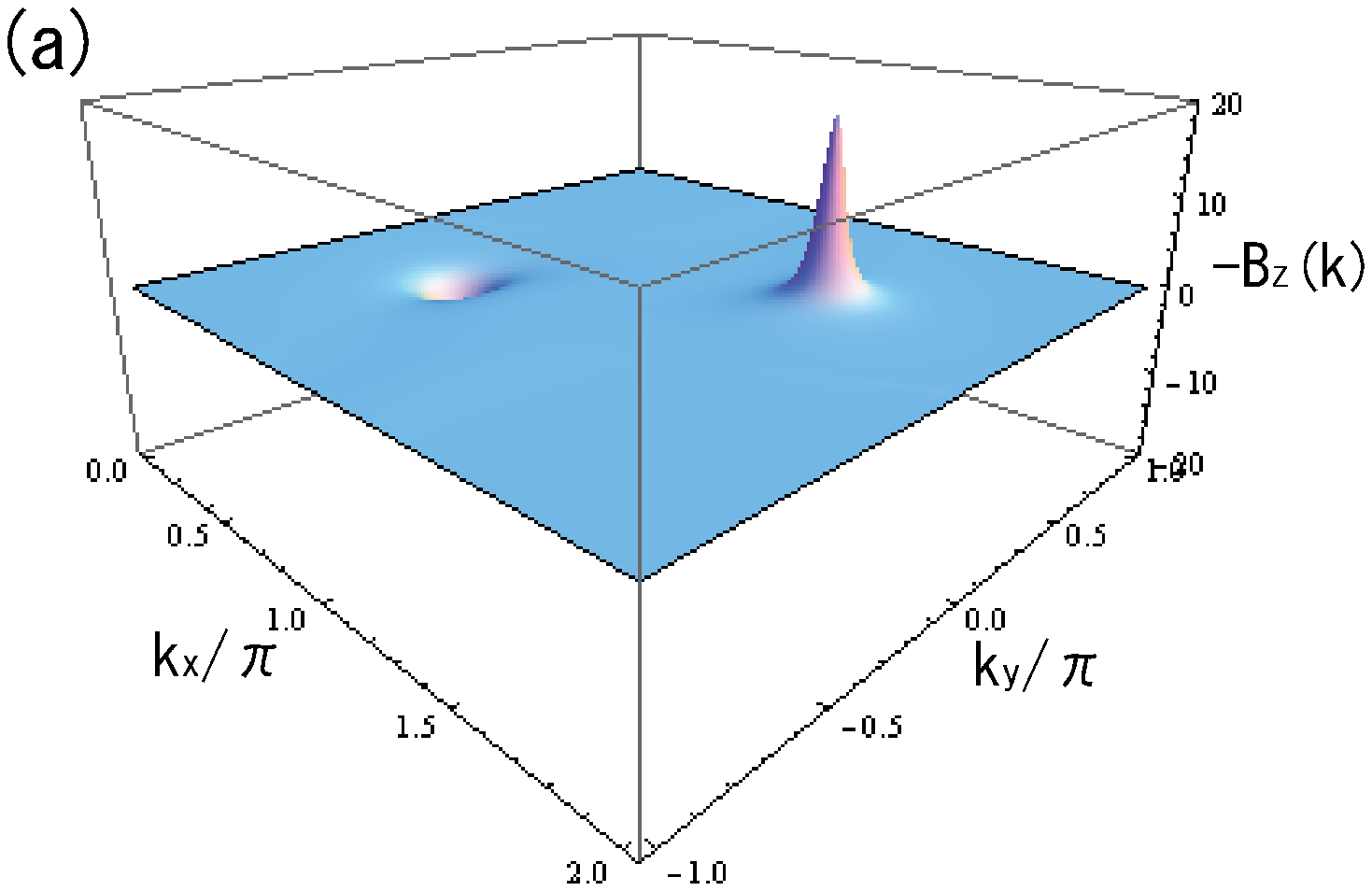}   
\includegraphics[width=6cm]{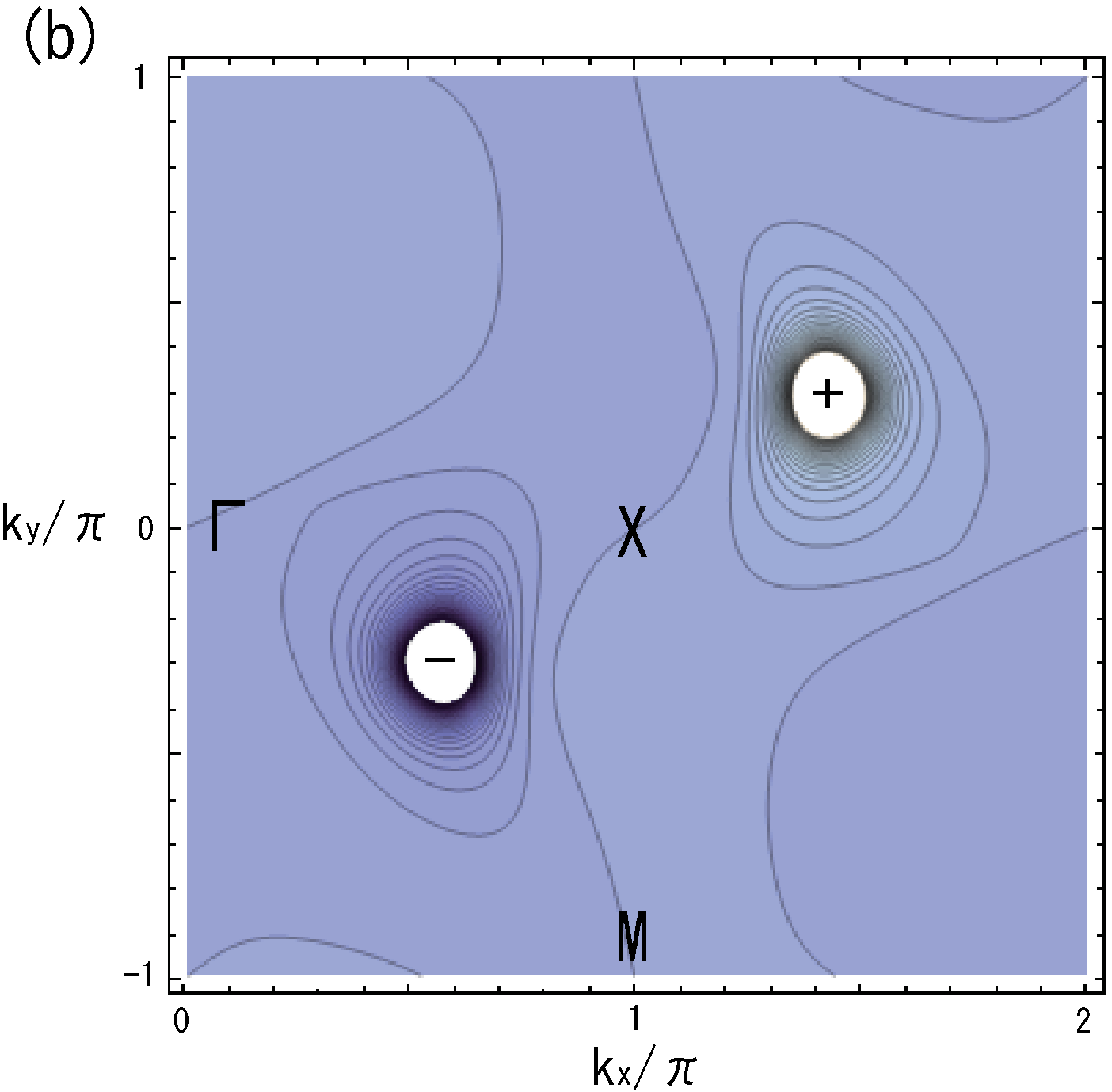}   
  \caption{(Color online)
Berry curvature and the contour plot of 
  $- \bm{B}\cdot \bm{e}_z$ for the energy band $E_1(\bm{k})$ 
          on the $k_x$-$k_y$ plane corresponding to  Fig. 1(a), 
where $\bm{e}_z$ denotes the unit vector perpendicular to $k_x$-$k_y$ plane. 
The middle of the horizontal axis (vertical axis) in the  contour plot 
   denotes the M ($ {\rm \Gamma}$)) point. 
}
\label{fig:fig2}
\end{figure}

The energy bands  $E_1$ and $E_2$  are  shown
  in Figs.~\ref{fig:fig1}(a) and \ref{fig:fig1}(b), respectively.
The Dirac cone exists at ${\bf k} \simeq \pm {\bf k}_0$ with
  ${\bf k}_0 = (0.57 \pi, - 0.3\pi)$, which is shown by an arrow. 
A pair of cones is seen around $E_1({\bf k}_0)$ and $E_2({\bf k}_0)$, 
 and the cones in the same band 
are symmetric with respect to the $\Gamma$(=(0,0)) point.
For $E_1(\bm{k})$,  the maximum is seen at the Y(=$(0, \pm \pi)$) point
 while saddle points are seen for the X(=$(\pi, 0)$),  M(=$(\pi, \pm \pi)$),
  and $\Gamma$ points.  
For $E_2(\bm{k})$,  the minimum is seen at the M point while saddle points are seen for the X, Y, and $\Gamma$ points.  
The accidental degeneracy, which is found at $\pm \bm{k}_0$ for $\Delta_0 \rightarrow 0$,
  is realized as the minimum of $E_1(\bm{k})$
and the maximum of $E_2(\bm{k})$.

   The Berry curvature, which is calculated using eq.~(\ref{eq:formula_2}), 
   is shown in Figs.~\ref{fig:fig2}(a) and \ref{fig:fig2}(b) for the three-dimensional 
(3D) view 
     on the $k_x$-$k_y$ plane (a), and for the contour plot (b).  
 The peak around $ - \bm{k}_0$ ( $\bm{k}_0$) is positive (negative). 
 They are antisymmetric with respect to the $\Gamma$ point
  (the middle point on the vertical axis).
Since the curvature exhibits a noticeable peak close to 
$\bm{k}_0$,   such a peak may be identified as the Dirac particle
 instead  of calculating the contact point. 
We note that 
 a quantity called a Chern number \cite{Chern} is given by 
\begin{align}
  \frac{1}{2\pi}\int_{S} {\rm d} {S} \; \bm{B} \cdot \bm{e}_z = 0,
 \end{align}
 owing to the time reversal symmetry, 
i.e., $\bm{B}(-\bm{k}) = - \bm{B}(\bm{k})$, 
 where 
 $S$ denotes the total BZ.
In Fig.~\ref{fig:fig2}(a),  the curvature around the Dirac point is slightly anisotropic
 in the sense that it is slightly extended to the $y$-direction.
With increasing pressure, the difference is reduced and the curvature becomes almost isotropic at $P_a$ = 10 kbar, where   two Dirac particles are 
well isolated.

\begin{figure}
  \centering
\includegraphics[width=6cm]{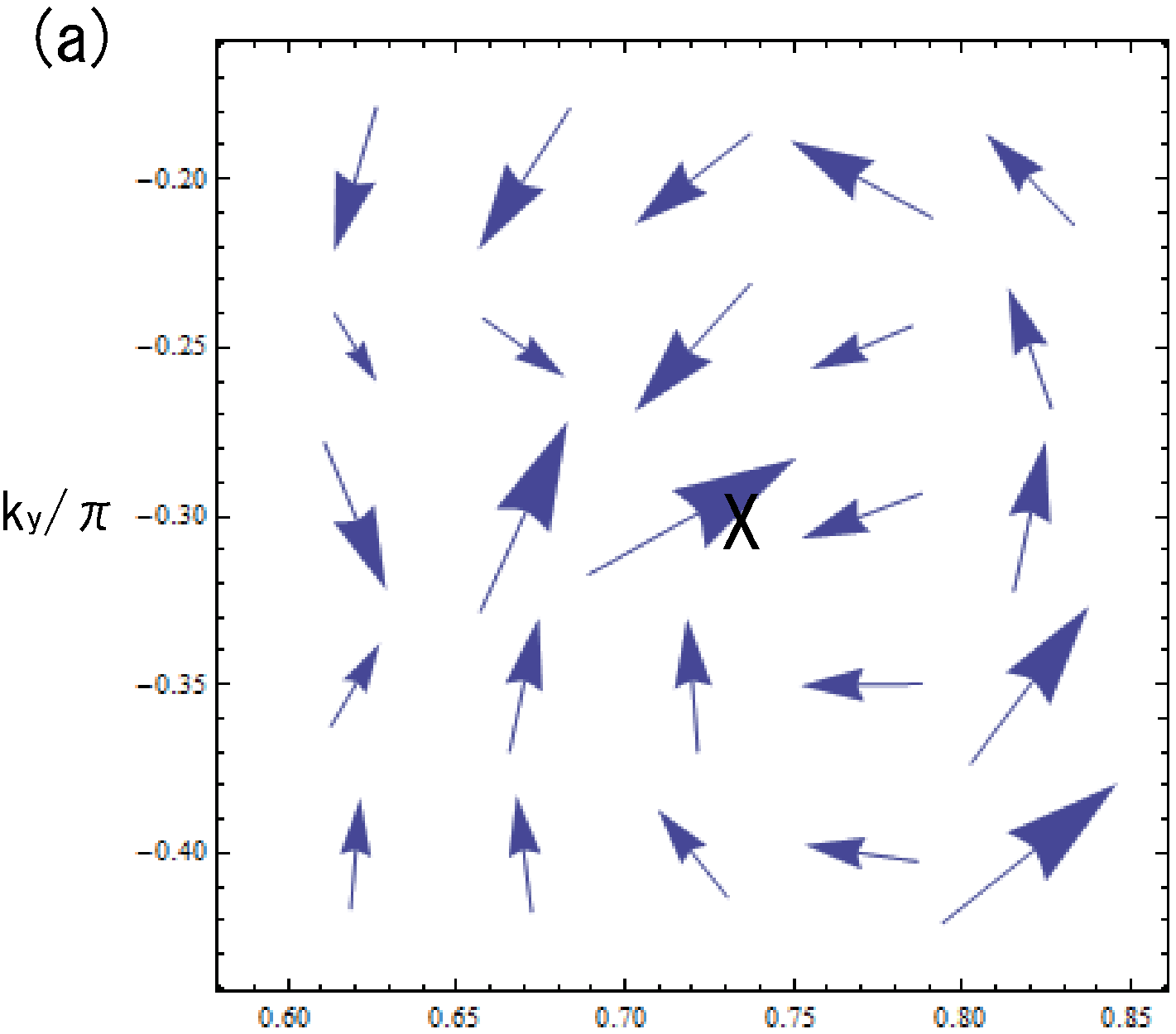}  
\includegraphics[width=6cm]{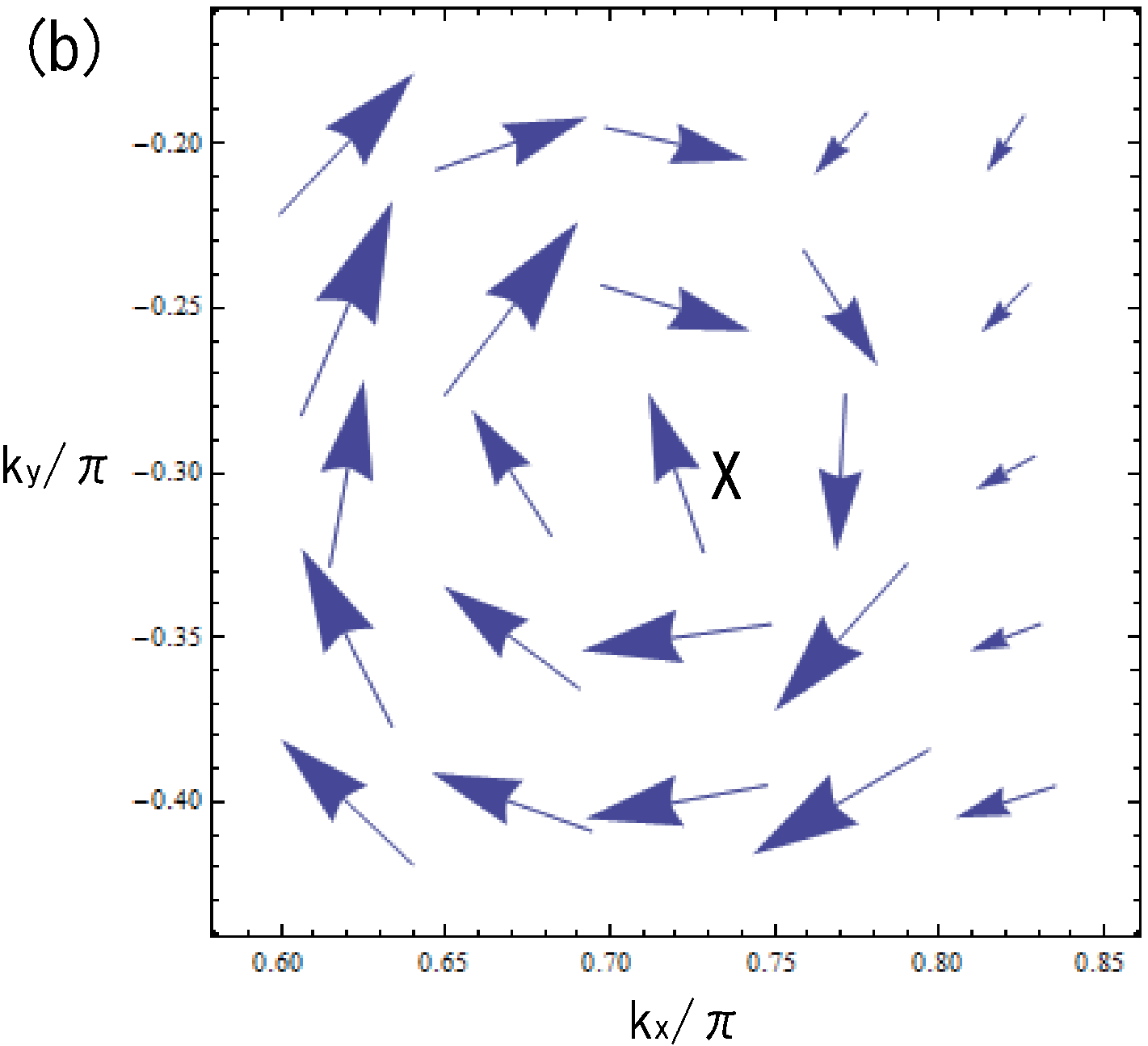}    
  \caption{(Color online)
Vector fields of velocities $ -\bm{v}_1$ (a) and 
   $ -\bm{v}_2$ (b) on the $k_x$-$k_y$ plane 
    around the Dirac point, $\bm{k}_0$,(cross)
     in Fig.~2.
}
\label{fig:fig3}
\end{figure}
Using eq.~(\ref{eq:def_velocity}), 
  we examine velocity field as a function of $\bm{k}$. 
In Figs.~\ref{fig:fig3}(a) and \ref{fig:fig3}(b), the $\bm{k}$ dependence of 
   the velocities $\bm{v}_1$ and $\bm{v}_2$ around $\bm{k}_0$ are shown.
Since the phases of both $|1(\bm{k})>$ and $|2({k})>$ are arbitrary, we choose them  so that  both of their first components corresponding to the A site are real.  
We verified that $\bm{B}$ calculated in terms of 
     $\bm{v}_1(\bm{k}) \times \bm{v}_2(\bm{k})$ 
    in eq.(\ref{eq:formula_reduced3}) reproduces well 
      the result in Fig.~\ref{fig:fig2} within numerical accuracy. 
 Both $\bm{v}_1$ and $\bm{v}_2$, which are almost orthogonal, rotate
      around the Dirac point,   indicating the singularity at  $\bm{k}_0$. 
The vortex structure around the Dirac point is also understood 
      from eqs.~(\ref{eq:eq41}) and (\ref{eq:eq42}). 
  Even in the presence of the site potential $\Delta_0$, 
    which removes the degeneracy of the zero-gap state, 
     the existence of such vortex behavior suggests 
       a topological property of the  Dirac particle.

\begin{figure}
  \centering
\includegraphics[width=6cm]{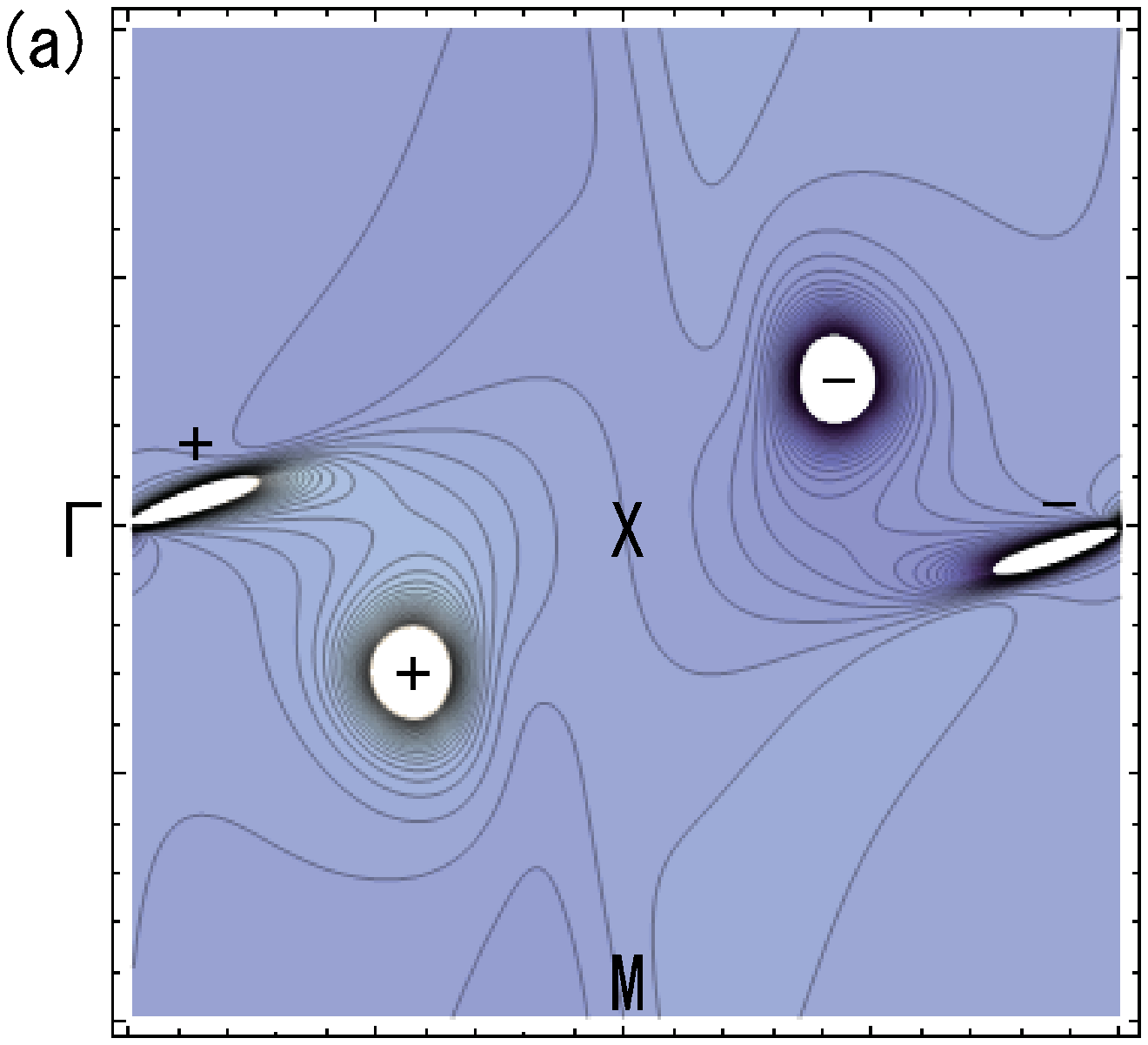}   
\includegraphics[width=6cm]{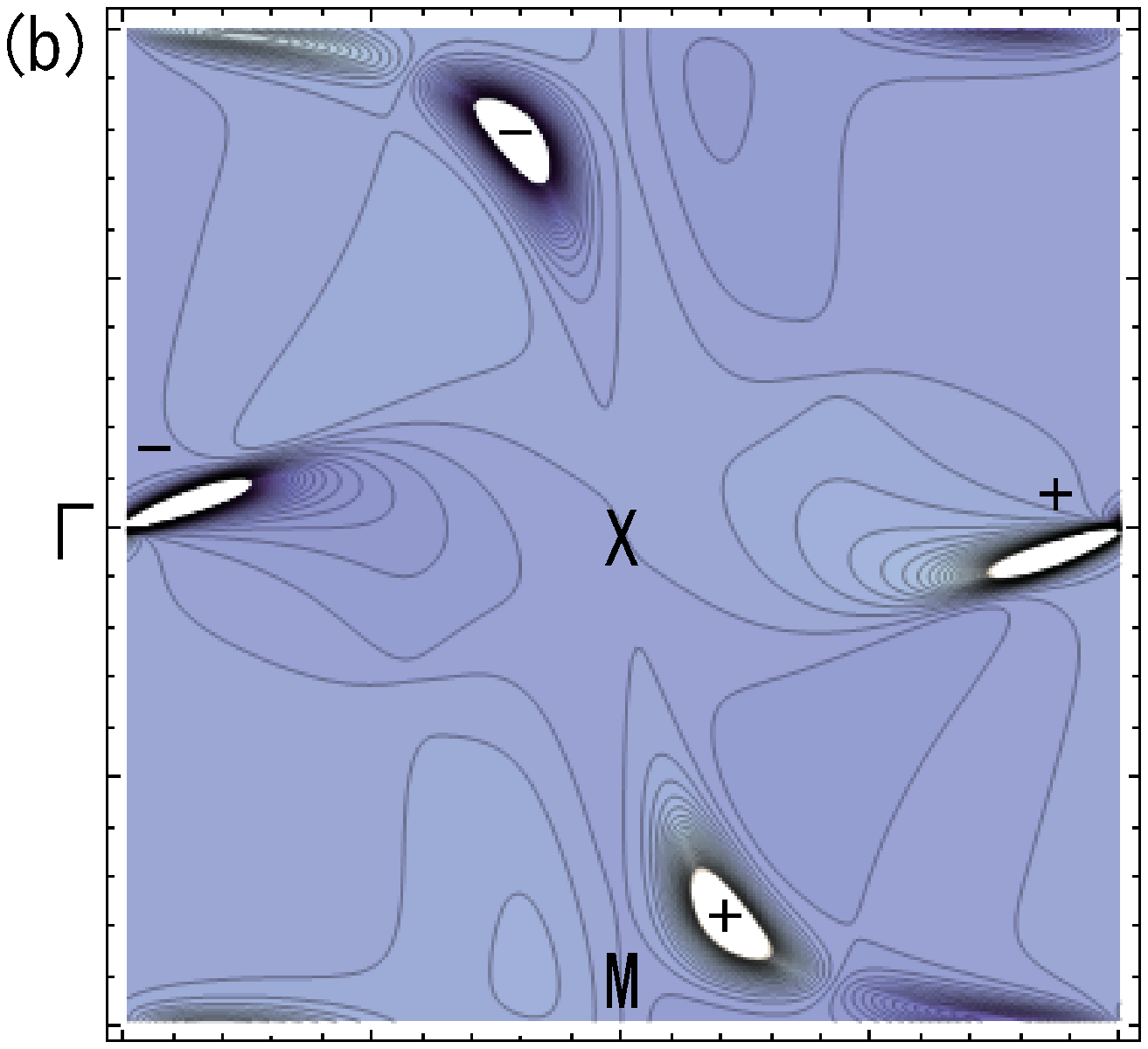}   
  \caption{(Color online)
Contour plot of Berry curvature, $B(\Delta_0)$,
 for  energy bands $E_2(\bm{k})$ (a)  and $E_3(\bm{k})$ (b). 
}
\label{fig:fig5m}
\end{figure}

Here, we examine the Berry curvature 
    for other bands of $E_2(\bm{k})$,  $E_3(\bm{k})$, and $E_4(\bm{k})$.
The Berry curvature of the second band $E_2(\bm{k})$ is shown 
    in Fig.~\ref{fig:fig5m}(a).
The peak located at $\pm \bm{k}_0$  has a sign 
    opposite to that of $E_1(\bm{k})$. 
In addition to such a peak, another pair of peaks appears close 
    to the $\Gamma$ point. 
 The latter one is rather extended to a direction slightly declined 
    toward the horizontal axis, suggesting a large anisotropy of the Dirac cone. 
The anisotropic peak disappears for a small $P_a <$ 3 kbar 
   while it becomes rather isotropic for a large $P_a$. 
 Such a behavior resembles the emergence of the Dirac particle 
    in the charge ordered  state, which has been shown   
      in $\alpha$-(BEDT-TTF)$_2$I$_3$.\cite{Kobayashi2010Merging} 
In Fig.~\ref{fig:fig5m}(b), the Berry curvature for $E_3(\bm{k})$
    is shown. 
 A pair of peaks close to the $\Gamma$ point also shows a sign
   opposite to that of $E_2(\bm{k})$. 
There is another peak  close to the M point, which  
   compensates for that of  $E_4(\bm{k})$ (not shown here). 
These results show that  each neighboring band provides 
    a pair of  Dirac particles  followed by the Berry curvature with an opposite sign. 
When one pair of Dirac particles  between neighboring bands is found,  
 one may expect another pair between  neighboring bands.

\begin{figure}
  \centering
\includegraphics[width=6cm]{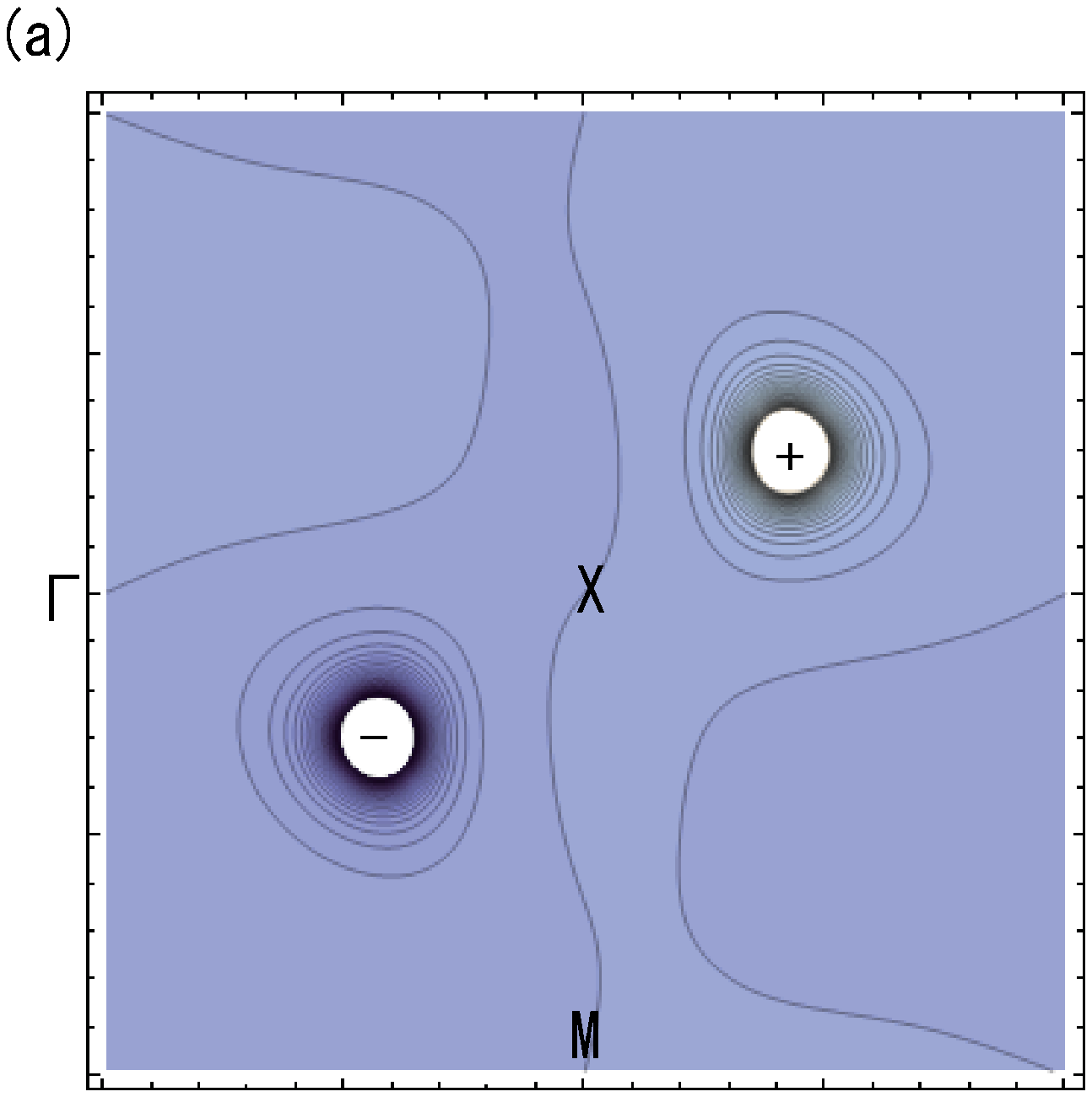}   
\includegraphics[width=6cm]{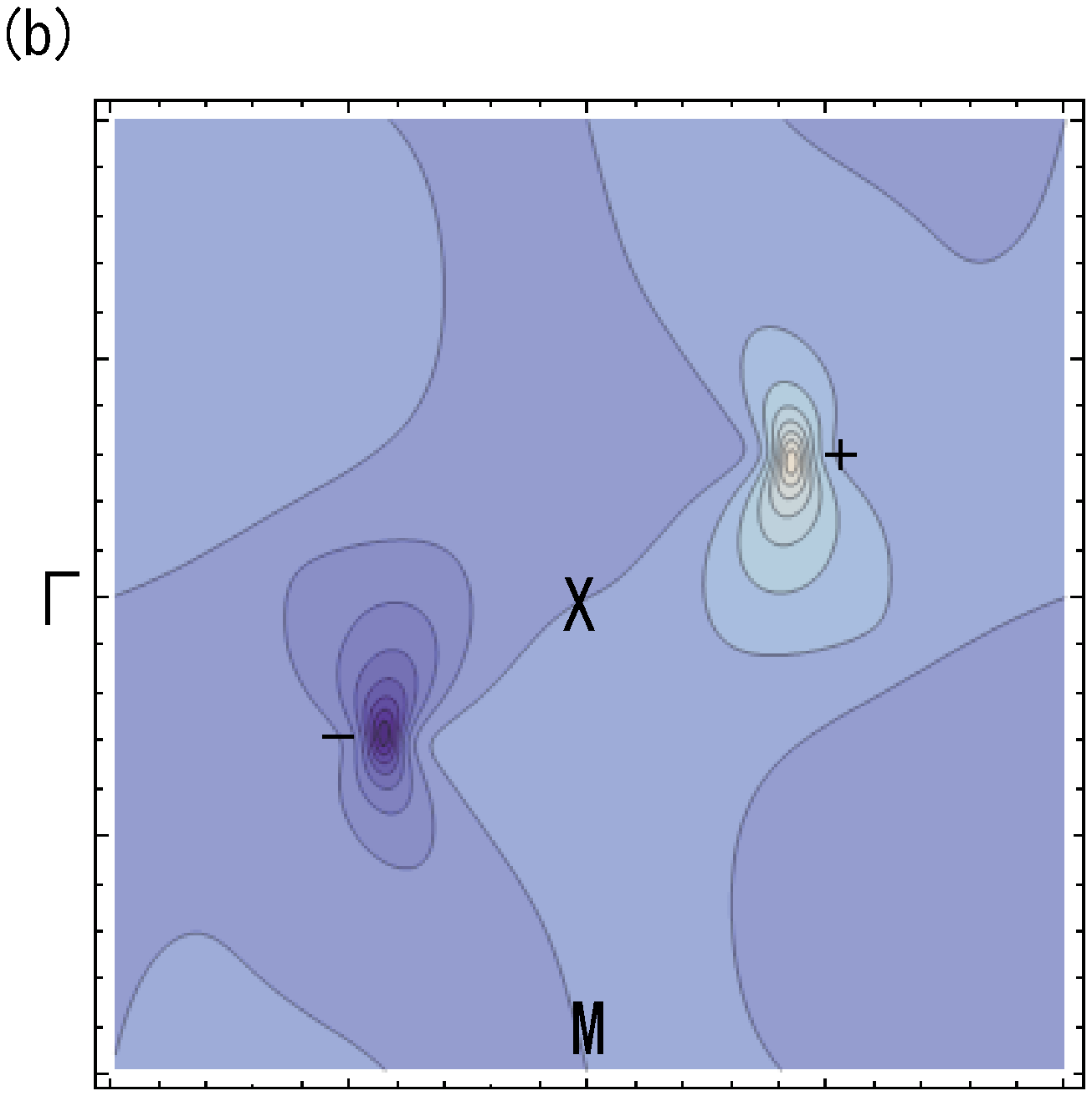}   
  \caption{(Color online)
Contour plots of the   Dirac component of eq.~(\ref{eq:B_Dirac}) (a),
 and 
  the background component of  eq.~(\ref{eq:B_back}) (b),
 for   Berry curvature of  $E_1(\bm{k})$
  in  Fig.~\ref{fig:fig2}. 
}
\label{fig:fig2A}
\end{figure}

Since the Berry curvature $\bm{B}(\bm{k})$ in Fig.~\ref{fig:fig2}  
 depends  on the sign of the site potential  $\pm \Delta_0$,   
  we examine $\bm{B}(\bm{k})$ by dividing it into 
     two components, i.e.,    $B_{\rm Dirac}$ and  $B_{\rm back}$, which are
   symmetric and  antisymmetric, respectively,  with respect to $\Delta_0$.   
By  defining  
\begin{eqnarray} 
 B(\Delta_0) = \bm{B}_1(\bm{k}, \Delta_0) \cdot \bm{e}_z  \; ,
\end{eqnarray}
these components are rewritten as  
\begin{align}
 B_{\rm Dirac} & = \frac{B(\Delta_0) - B(-\Delta_0)}{2} \; ,
\label{eq:B_Dirac} \\
 B_{\rm back} & = \frac{B(\Delta_0) + B(-\Delta_0)}{2} \; .
\label{eq:B_back}
 \end{align}
The first one is related to the topological property, 
   which does not depend on the details of the energy band. 
The second one depends on the property of the energy band.  
The curvatures given by  eqs.~(\ref{eq:B_Dirac}) and (\ref{eq:B_back}) 
    are shown in Figs.~\ref{fig:fig2A}(a) and \ref{fig:fig2A}(b),
 respectively. 
The peak height of $B_{\rm Dirac}$ is  $10^2$ times as much as 
 that of   $B_{\rm back}$. 
 When $\Delta_0$ decreases to zero, 
  $B_{\rm Dirac}$ increases rapidly, but 
     $B_{\rm back}$ remains almost unchanged. 
The component  $B_{\rm Dirac}$   is located 
 in the narrow region around  $\pm \bm{k}_0$, 
  suggesting  a  character of the Dirac particle.
The component $B_{\rm back}$, which  is small and  spreads in the entire region of the BZ,  is irrelevant to the Dirac particle.   Thus, the separation of 
 $B_{\rm Dirac}$ from $B(\Delta_0)$  gives  the intrinsic curvature 
 of the Dirac particle.

\begin{figure}
  \centering
\includegraphics[width=8cm]{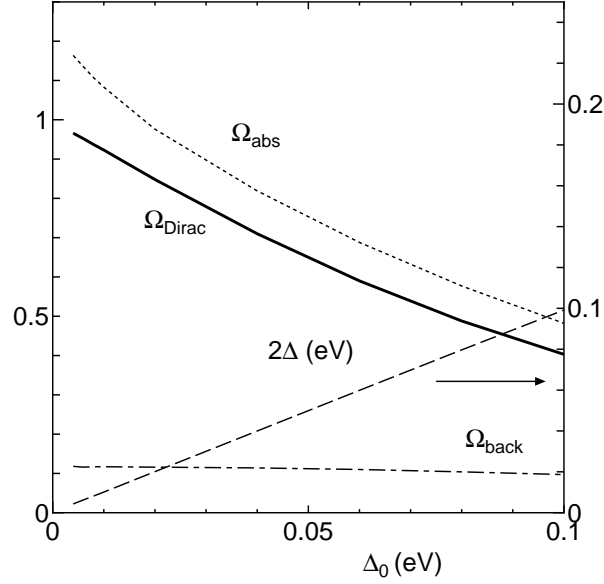}   
  \caption{
  $\Delta_0$ dependence of 
       $\Omega_{\rm Dirac}$ (eq.~(\ref{eq:phase_1})), 
       $\Omega_{\rm back}$, (eq.~(\ref{eq:phase_2})), 
       $\Omega{\rm abs}$ (eq.~(\ref{eq:phase_3})), 
   and  gap $2 \Delta$ . 
}
\label{fig:Berry_CO}
\end{figure}

To estimate the magnitude of the peak of $B(\bm{k}, \Delta_0)$,
 we examine the following integrated quantities of the respective components: 
\begin{align}
 \Omega_{\rm Dirac} &= \frac{1}{2\pi}\int_{S} {\rm d} {S} \; 
 {\rm sgn}(k_y) \frac{B(\Delta_0) - B(-\Delta_0)}{2} \; ,
\label{eq:phase_1}
\\ 
 \Omega_{\rm back} &= \frac{1}{2\pi}\int_{S} {\rm d} {S} \;  
  {\rm sgn}(k_y) \frac{B(\Delta_0)+ B(-\Delta_0)}{2} \; .
\label{eq:phase_2} 
 \end{align}
The quantity  $\Omega_{\rm Dirac}+\Omega_{\rm back}$  is a Berry phase 
 where   $S$ in eq.~(\ref{eq:Berry_C}) is  the region given by   
 $- \pi < k_x < \pi$ and 
$0 < k_y < \pi$.
The quantity  $\Omega_{\rm Dirac}$ comes from the  contribution of the Dirac cone 
 with a singularity, and the second one originates 
   from the property of the conventional band.  
Note that  $\Omega_{\rm Dirac}$ remains the same but 
    $\Omega_{\rm back}$ varies when we replace 
     ${\rm sgn}(k_y)$ by  ${\rm sgn}(k_x)$ in the limit of small $\Delta_0$. 
In Fig.~\ref{fig:Berry_CO}, 
  $\Omega_{\rm Dirac}$ and  $\Omega_{\rm back}$ are shown as the function of 
     $\Delta_0$. 
With decreasing $\Delta_0$,   $\Omega_{\rm Dirac}$ is reduced to 1,
  since  the boundary effect of BZ becomes negligible.   
This indicates the intrinsic property of Dirac cones.
 We also show the quantity
\begin{align}
\Omega_{\rm abs} &= \frac{1}{2\pi}\int_{S} {\rm d} {S} \; 
  \frac{|B(\Delta_0)| + |B(-\Delta_0)|}{2} .
\label{eq:phase_3}
 \end{align}
Note that $\Omega_{\rm abs}$ is larger than  $\Omega_{\rm Dirac}$ 
 and $\Omega_{\rm back}$, and becomes even larger than 1 for a small $\Delta_0$.
The enhancement of $\Omega_{\rm abs}$ is attributable to 
 the alternation of the sign of $B(\Delta_0)$ 
  as a function of ${\bf k}$, which  
   originates from the background  property of the band structure.

\section{Summary and discussion}
We obtained the following  for the Berry curvature of  $\alpha$-(BEDT-TTF)$_2$I$_3$, which exhibits the zero-gap state under the uniaxial pressure. 

(i)~In the present study, instead of  the zero gap between  $E_1(\bm{k})$ and $E_2(\bm{k})$, the Dirac particle is examined  by calculating the Berry curvature for the 4x4 Hamiltonian. 
By adding a small potential $\Delta_0$ acting  on the A and A' sites 
  with  opposite signs, which  breaks the inversion symmetry,
 the Dirac particle is identified  by  the pronounced peak of  
   the  Berry curvature in the same band,
  and also between  neighboring bands. 
The Berry curvature consists of two components,i.e.,  $B_{\rm Dirac}$ and $B_{\rm back}$, where   $B_{\rm Dirac}$ ($B_{\rm back}$) is an odd (even) function with respect to $\Delta_0$.  
The peak in $B_{\rm Dirac}$, which is  located close to the Dirac cone,  
 is intrinsic to the Dirac particle. 
 The quantity $B_{\rm back}$, which  
  is small but is extended  in the whole BZ, is irrelevant to the Dirac particle. 
The latter could be the origin of the curvature, which depends 
on  the choice of the phase in the transfer integrals  of the site Hamiltonian.
\cite{Bena2009}

(ii)~The Berry curvature is calculated in the general case of the reduced 2x2 Hamiltonian. 
Once  such  Hamiltonian (i.e., the coefficient of Pauli matrices) 
   for other organic conductors with possible  Dirac particles is obtained  on the L-K basis, 
   eq.~(\ref{eq:Berry_curvarure0}) is applicable  to understand 
    the characteristic behavior of the Berry curvature. 
The Berry curvature is determined by a vector product of two types of  velocity fields, which are given as the coefficients of the Pauli 
 matrices, $\sigma_1$ and $\sigma_2$, in the reduced Hamiltonian.  
 These velocity fields  rotate around the Dirac point as a vortex and 
 are  determined uniquely, except for the Dirac point. 

(iii)~We found a pair of Dirac particles  not only between $E_1(\bm{k})$ and $E_2(\bm{k})$ , but also between the other neighboring band, although the mutual relation  of these  Dirac points  is unclear owing to 
the accidental degeneracy. In the present case, the curvature 
 between $E_2(\bm{k})$ and $E_3(\bm{k})$ exhibits a behavior close to  merging. 
It is possible that  a pair of Dirac particles   may exist  in many  neighboring bands, if a pair of peaks is found  in one band.

Here, we  comment on the effect of a short-range repulsive interaction
 on the Dirac particle.  
In the presence of the interaction that also  gives  the normal state, i.e., no gap,  the critical pressure for the zero-gap state ($P_a \sim 4$ kbar)
\cite{Kobayashi2007JPSJ} 
 is larger than that in the absence of interaction ($P_a \sim 3$ kbar), 
 suggesting that the interaction suppresses the zero-gap state. 
 Such an effect appears as the on-site potential acting on the B and C sites.
Comparing the Berry curvature in Fig. \ref{fig:fig2}(a) with  that in the presence of the interaction (not shown here), we found that the Berry curvature in the presence of the interaction becomes  rather anisotropic and  exhibits a small tail toward the M-point, indicating a tendency toward merging. 
Thus,  merging may occur owing to the  interaction,
 in addition to the variation in the transfer energy.
 \cite{Katayama2006JPSJ}  
 Note that the pair of  Dirac
particles already merges  at the M-point in the charge ordering state at ambient pressure, which was  obtained in our previous study.\cite{Kobayashi2005JPSJ}

At present, it is not yet clear how   Dirac particles emerge 
 in organic conductors, since it occurs accidentally in the BZ. 
 However, the calculation of the  Berry curvature is a useful tool 
   for searching  Dirac particles particularly in the multiband system 
 even when the symmetry is broken or when the band structure is complicated. 
Actually, it is shown that 
 the organic conductor $\alpha$-(BEDT-TTF)$_2$NH$_4$Hg(SCN)$_4$ 
 with  a complicated band structure 
  exhibits  Dirac particles with a zero gap under pressure.\cite{Choji} 
 In the case where the symmetry is broken, 
      the formation of massive Dirac particles in the  stripe charge ordered state        is reported in a separate paper.
\cite{Kobayashi2010Merging}

\acknowledgements
The authors are thankful to 
F. Pi\'echon, J.-N. Fuchs, and G. Montambaux
 for useful discussions in the early stage of the present work. 
Y.S. is indebted to the Daiko Foundation for financial aid
 in the present work.
This work was financially supported in part
 by a Grant-in-Aid for Special Coordination Funds for Promoting
Science and Technology (SCF) and for Scientific Research on Innovative
Areas 20110002, 
and by Scientific Research (Nos. 19740205, 22540366, and 23540403)
 from the Ministry of
Education, Culture, Sports, Science and Technology in Japan.


\appendix

\section{Berry curvature and velocity field  for general Hamiltonian} 
We calculate the Berry curvature for the reduced 2x2  Hamiltonian 
 given by  (eq.~(\ref{eq:Heff_start})), 
 \begin{align}
 H^{\rm LK}(\bm{k}) 
 & = 
 \begin{pmatrix}
f_3 & f_1-if_2 \\
f_1+if_2 & - f_3 
\end{pmatrix} ,
\label{eq:Heff_start1}
\end{align}
 where the quantity $\sigma_j$ denotes the Pauli matrix
 and the energy is measured from the chemical potential.
The quantity $f_j$ (= $f_j(\bm{k})$)   depends on
  the two-dimensional wave vector $\bm{k} (=  \bm{k}_0 + \bm{q})$. 
The Schr\"{o}dinger equation for the  Hamiltonian (\ref{eq:Heff_start1})  is written as  
\begin{align}
H^{\rm red} \Psi_{\pm} = \pm E \Psi_{\pm},
\label{eq:eq17}
\end{align}
where $\Psi_+ = |1(\bm{k})>_0$, $\Psi_-= |2(\bm{k})>_0$.
The energy and wave function are calculated as 
\begin{align}
 & E = \sqrt{f_1^2 + f_2^2 + f_3^2} \; ,
\label{eq:eq18}
  \\
 &\Psi_{\pm} = \frac{1}{\sqrt{2E(E \mp f_3)}} 
 \begin{pmatrix}
  f_1-if_2 \\
   \pm E  - f_3
\end{pmatrix} 
 \equiv 
\begin{pmatrix}
  \tilde{f}_1(\bm{k})-i\tilde{f}_2(\bm{k}) \\
  g_{\pm}(\bm{k})
\end{pmatrix}
\label{eq:eq19}
.
\end{align} 
The relative phase between  $\Psi_+$ and $\Psi_-$  
 is undetermined and is chosen such that
  the second component of each state is  real and positive.  

Now, we calculate $\bm{B}_1$  for $\Psi_+$  explicitly 
using  eq.~(\ref{eq:formula_1}). 
By noting that only the first component  contains 
the imaginary part, 
 the Berry curvature $\bm{B}(\bm{k})$ is calculated  as 
\begin{align} 
 \bm{B_1}& = \bm{B} = -  {\rm Im} \left( \bm{\nabla}_{\bm{k}}\Psi_+^{\dagger} \right)^t 
  \times \left( \bm{\nabla}_{\bm{k}} \Psi_+  \right)
 \nonumber \\ 
 & = - {\rm Im} (\bm{\nabla}_{\bm{k}}\tilde{f}_1+ i \bm{\nabla}_{\bm{k}}\tilde{f}_2) \times 
  (\bm{\nabla}_{\bm{k}}\tilde{f}_1- i \bm{\nabla}_{\bm{k}}\tilde{f}_2) 
 = 2 \bm{\nabla}_{\bm{k}} \tilde{f}_1 \times \bm{\nabla}_{\bm{k}} \tilde{f}_2 \; ,
\label{eq:cal1}
\end{align} 
where 
\begin{align} 
  \bm{\nabla}_{\bm{k}}\tilde{f}_1 = & \frac{1}{\sqrt{2E(E-f_3)}} \{  
        ( 1 - \frac{(2E-f_3)f_1^2}{2E^2(E-f_3)} \bm{\nabla}_{\bm{k}} f_1 
     \nonumber \\
     &    - \frac{(2E-f_3)f_2f_1}{2E^2(E-f_3)} \bm{\nabla}_{\bm{k}}f_2 
       + \frac{(E-f_3)f_1}{2E^2} \bm{\nabla}_{\bm{k}} f_3 \} \; ,
\label{eq:cal2}
  \end{align} 
 and 
\begin{align} 
  \bm{\nabla}_{\bm{k}}\tilde{f}_2 = & \frac{1}{\sqrt{2E(E-f_3)}} \{ 
         - \frac{(2E-f_3)f_2f_1}{2E^2(E-f_3)} \bm{\nabla}_{\bm{k}}f_1 
     \nonumber \\
   &   +  ( 1 - \frac{(2E-f_3)f_2^2}{2E^2(E-f_3)} \bm{\nabla}_{\bm{k}}f_2 
       + \frac{(E-f_3)f_2}{2E^2} \bm{\nabla}_{\bm{k}} f_3 \}.
\label{eq:cal3}
  \end{align} 
Substituting eqs.~(\ref{eq:cal2}) and (\ref{eq:cal3}) into 
 eq.~(\ref{eq:cal1}), we obtain  eq.~(\ref{eq:Berry_curvarure0}).

 Note that  eq.~(\ref{eq:Berry_curvarure0})  can  also be 
 obtained  from  eq.~(\ref{eq:formula_2}), which is rewritten as  
\begin{align}
\bm{B} & = - {\rm Im}  \left\{ 
 ( \Psi_+^{\dagger} \bm{\nabla}_{\bm{k}} H \Psi_- ) \times 
  ( \Psi_-^{\dagger} \bm{\nabla}_{\bm{k}} H \Psi_+ ) \frac{1}{4E^2} 
 \right\}
  = 
- \frac{1}{4E^2}
 \times 
 \nonumber \\
& {\rm Im} 
 \left\{ \right.
 (\Psi_+| \sigma_1 \bm{\nabla}_{\bm{k}}f_1 + \sigma_2 \bm{\nabla}_{\bm{k}}f_2 
   + \sigma_3 \bm{\nabla}_{\bm{k}}f_3 |\Psi_-) 
 \times 
  (\Psi_-| \sigma_1 \bm{\nabla}_{\bm{k}} f_1 + \sigma_2 \bm{\nabla}_{\bm{k}}f_2 
    + \sigma_3 \bm{\nabla}_{\bm{k}}f_3 |\Psi_+) \left. \right\} ,
\label{eq:A10}
 \end{align} 
where $(\Psi_\pm| =( \tilde{f}_1 + i \tilde{f}_2 , g_\pm)^t$.
 Equation~(\ref{eq:A10}) is calculated as  eq.~(\ref{eq:Berry_curvarure0})
by noting  
$(\Psi_-|\sigma_1|\Psi_+)= 
- (f_3f_1 -  {\rm i} Ef_2)/
 (E \sqrt{E^2-f_3^2})
$, 
$(\Psi_-|\sigma_2|\Psi_+)
 = - (f_3f_2 +  {\rm i} Ef_1)/(E \sqrt{E^2-f_3^2})$,  
and 
$(\Psi_-|\sigma_3|\Psi_+)
 =  \sqrt{E^2-f_3^2}/E$. 

Here, we calculate the velocity field defined by eq.~(\ref{eq:def_velocity}).
 Substituting eq.~(\ref{eq:eq19}) into   eq.~(\ref{eq:def_velocity}), we obtain
 \begin{align}
\label{eq:v1_eff}
  \bm{v}_1(\theta_{\bm{k}}) 
= &  {\rm Re} \{ \Psi^\dagger_- \bm{\nabla}_{\bm{k}} H^{\rm LK} \Psi_+ \} 
 \nonumber \\
& =  
\frac{-1}{E \sqrt{E^2 -f_3^2}}
 \Large(
 f_3 (f_1 \bm{\nabla}_{\bm{k}}f_1 + f_2 \bm{\nabla}_{\bm{k}}f_2) 
 \nonumber  \\
 & - (f_1^2 + f_2^2) \bm{\nabla}_{\bm{k}}f_3 
\Large) ,
 \\ 
\label{eq:v2_eff}
 \bm{v}_2(\theta_{\bm{k}}) = 
&  {\rm Im} \{ \Psi_- \bm{\nabla}_{\bm{k}} H^{\rm LK} \Psi_+ \} 
 \nonumber \\ 
 & = 
 \frac{-1}{\sqrt{E^2 -f_3^2}}
 \Large(- f_2 \bm{\nabla}_{\bm{k}}f_1 + f_1 \bm{\nabla}_{\bm{k}}f_2 \Large)  \; ,
\\
\label{eq:v3_eff}
  \bm{v}_3(\theta_{\bm{k}}) = 
& \frac{1}{2} \{ 
\Psi^\dagger_+ \bm{\nabla}_{\bm{k}} H^{\rm LK} \Psi_+ 
 -  \Psi^\dagger_- \bm{\nabla}_{\bm{k}} H^{\rm LK} \Psi_-
 \} 
 \nonumber \\ 
 & =  \frac{1}{\sqrt{E^2-f_3^2}}
  \Large( f_1 \bm{\nabla}_{\bm{k}}f_1 + f_2 \bm{\nabla}_{\bm{k}}f_2 + f_3 \bm{\nabla}_{\bm{k}}f_3 
 \Large)  \; .
\end{align}
 Substituting eqs.~(\ref{eq:v1_eff}) and  (\ref{eq:v2_eff})
 into eq.~(\ref{eq:formula_reduced3}), one finds that 
$\bm{B}_1$ is exactly the same as eq.(\ref{eq:Berry_curvarure0}).

\end{document}